\documentclass[conference]{IEEEtran}
\IEEEoverridecommandlockouts
\usepackage{cite}
\usepackage[top=0.7in, bottom=1in, left=1in, right=1in]{geometry}

\usepackage{amsmath,amssymb,amsfonts}
\usepackage{algorithm}
\usepackage{algorithmicx}
\usepackage{algpseudocode}
\usepackage{graphicx,caption,subcaption}
\usepackage{textcomp}
\usepackage{xcolor}
\def\BibTeX{{\rm B\kern-.05em{\sc i\kern-.025em b}\kern-.08em
    T\kern-.1667em\lower.7ex\hbox{E}\kern-.125emX}}
\begin{document}

\title{Partially Reflected Surface (PRS)-Loaded Graphene-Based
Patch Antenna for 6G

}
\author{Omar Osman, Abdullah Qayyum,~and Maziar Nekovee \\
\textit{6G Lab, School of Engineering and Informatics}\\
\textit{University of Sussex,}, United Kingdom \\
Email:\{oo313,~A.qayyum,~M.nekovee\}@sussex.ac.uk
}

\maketitle

\begin{abstract}
This work investigates a slotted patch antenna integrated with a partially reflected surface (PRS) to operate in the TeraHertz (THz) frequency range for 6G. The antenna is based on graphene material, 
on a Rogers RT Duroid 6010 substrate. 
The proposed antenna achieves a bandwidth of 70 GHz (750 GHz to 820 GHz). The PRS sheet consists of 5x4 unit cells, which are optimised to enhance the overall realized gain of the antenna. The overall realized gain has increased by 1.07 dBi. 
Also, the PRS enhanced the antenna radiation pattern, showing stable properties over the
operating bandwidth. The improved antenna performance is validated via simulations.
\end{abstract}


\section{Introduction}
Sixth-generation (6G) wireless communication will redefine wireless connectivity by delivering ultra-high data rates, ultra-low latency, and massive connectivity for emerging applications such as holographic communications, immersive augmented reality, virtual reality, and intelligent transportation systems \cite{siddiky2025comprehensive}. To meet these requirements, the terahertz (THz) frequency band (0.1THz--10THz) has attracted considerable interest due to the vast available spectrum that can support data rates on the order of terabits per second \cite{chen2019survey}. However, practical THz communication systems must overcome substantial challenges, including severe free-space path loss, high atmospheric absorption, and device limitations \cite{shafie2022terahertz}.

Graphene, a two-dimensional material with exceptional electrical and mechanical properties, has emerged as a promising candidate for THz applications. Its high electron mobility, tunability, and capability to support surface plasmon polaritons enable the design of miniaturized, wideband antennas suitable for integration into next-generation devices \cite{abadal2022graphene}. Recent studies have demonstrated that graphene-based patch antennas can achieve broad operating bandwidths while maintaining a compact footprint, which is essential for modern wireless communication systems \cite{dash2022ultra, fakharian2022graphene, fakharian2025reconfigurable}.

Patch antennas consist of a dielectric substrate layer, a ground plane, and a patch as the top layer. The ground plane
and the patch are typically metallic. There are different geometries of patch antennas, which can yield different
radiation characteristics \cite{james1986microstrip}. However, the most commonly used are rectangular and circular patch antennas \cite{singh2011micro}. Patch antennas can be easily fabricated on a printed circuit board (PCB), which reduces complexities. They are
advantageous because they are low-cost and lightweight, in addition to their low-profile characteristics, making them easy to use \cite{omoleye2025development}. Patch antennas are used in aviation, automotive, wearable devices, mobile communications, and radar applications \cite{singh2011micro, pushpanjali2024literature, qayyum2020novel, dhadwal2025graphene, mujawar2023thz}. Due to the small size of patch antennas, these antennas are an ideal choice for mobile communication as the mobile handheld devices are small in size \cite{fares2010mobile}. Despite these advantages, planar patch antennas often suffer from low gain and narrow beamwidth. Various techniques, such as defected ground structure (DGS), photonic Bandgap (PBG), Electromagnetic Bandgap (EBG), and Partially Reflecting Surfaces (PRS), have been explored to address these limitations \cite{pant2021gain}. Implementing Partially Reflecting Surfaces (PRS) is emerging as one of the most effective strategies. A PRS consist of an array of unit cells. This sheet of unit cells is configured to
behave as a partially reflecting superstrate positioned above the antenna through a ground structure \cite{trentini1956partially}. A PRS can enhance antenna performance by inducing a leaky-wave mechanism that creates multiple reflections between the antenna and the PRS. This process results in constructive interference and beam collimation, improving the realized gain without significantly increasing the antenna profile. In \cite{9618202}, adding a phase-compensated metasurface above the PRS led to a peak gain of approximately 24.9 dBi, with a 1 dB gain bandwidth of 5\%. A three-layer PRS achieved a 3 dB gain bandwidth of 15\% and a maximum gain of 20 dBi at 14.5 GHz \cite{zhu2022broadband}.

This work proposes a PRS-loaded graphene patch antenna for 6G THz applications. The proposed antenna integrates a wideband slotted graphene patch with a carefully optimized PRS array to enhance its gain and stabilize its radiation pattern over a 70 GHz bandwidth. Simulation results indicate that the PRS not only shifts the resonance frequency but also increases the peak realized gain by over 1 dBi. The remainder of the paper is organized as follows: Section II details the antenna and PRS design, Section III presents the simulation setup and results, and Section IV concludes the paper with an outlook on future work.

\section{Proposed Antenna}

\subsection{Patch Antenna}

\begin{figure}  
		\centering
		\subfloat[Front View ]{\includegraphics[width=0.28\textwidth]{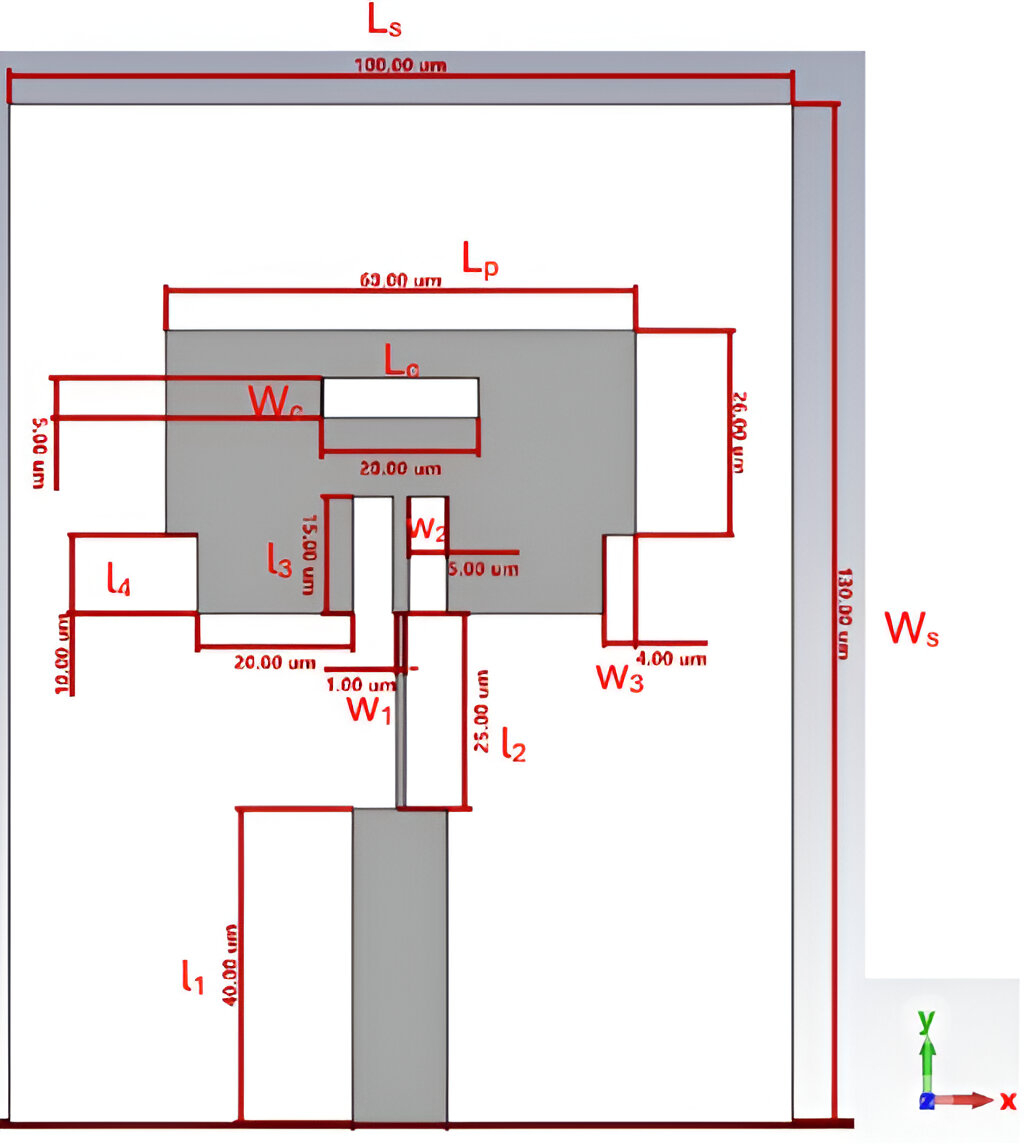} }\quad
		\subfloat[Back view]{\includegraphics[width=0.25\textwidth]{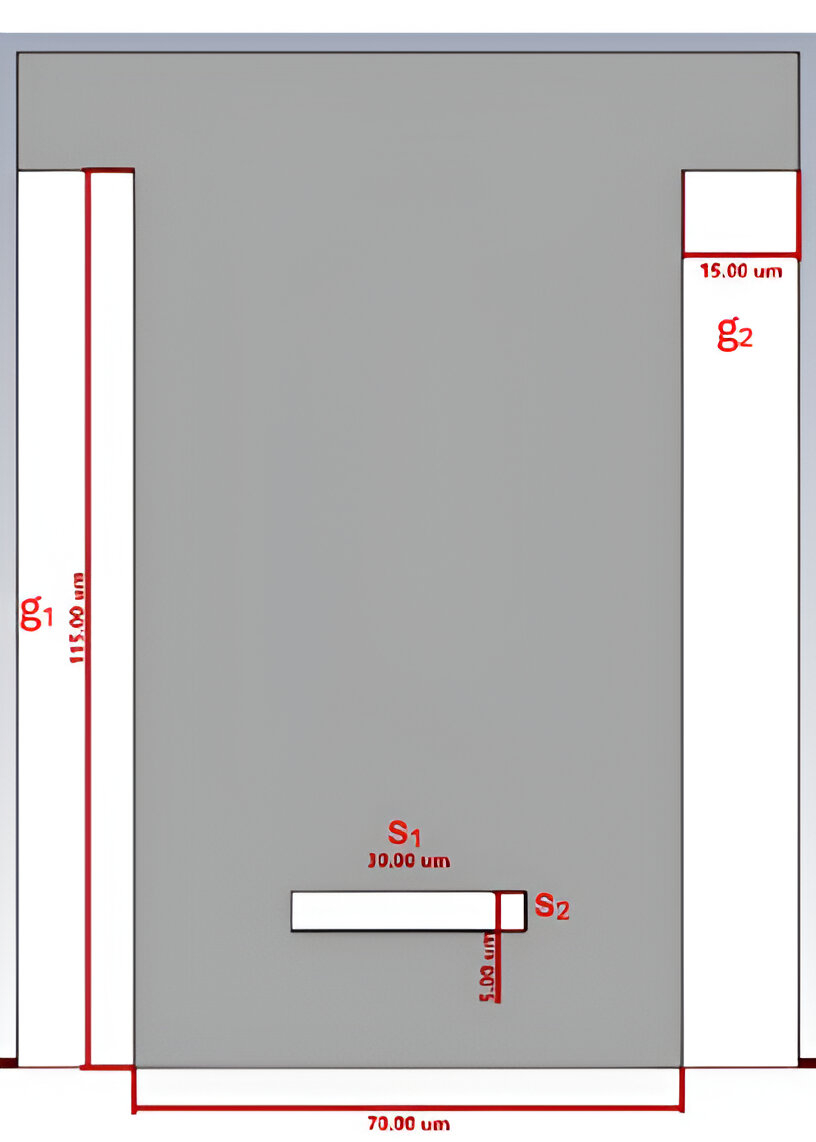}  }
		\caption{Geometry of Patch Antenna.}
		\label{fig:1}
	\end{figure}

The proposed patch antenna geometry with DGS is shown in Fig. \ref{fig:1}. The given antenna is designed on Rogers (RT6010) substrate with dielectric constant \begin{math}(\varepsilon_r)\end{math}  10.2 and loss tangent \begin{math}(tan \delta )\end{math} 0.0023 thickness 0.45$\mu $m with the radiating element made up of graphene. Computer Simulation Tool (CSTMWS 2024) is used to design and simulate the antenna.

As shown in Fig. \ref{fig:1}(b), a ground plane made of graphene with a DGS is placed below the substrate. The DGS is introduced to improve the impedance matching accuracy and increase the antenna's bandwidth. The proposed antenna is designed to have a resonance frequency of 800 GHz. The following expression calculates the width of the patch:

\begin{equation}
W_p=\frac{c}{2 \mathrm{f}_r} \sqrt{\frac{2}{\varepsilon_r+1}},
\end{equation}
where $c$ is the speed of light, $f_r$ is the resonance frequency and $\varepsilon$ is the relative permittivity.

\begin{table}
\center
\setlength{\tabcolsep}{1.2em}
\renewcommand{\arraystretch}{1.8}
\caption{Optimized parameter values of Patch Antenna} \label{tab1}
\begin{tabular}{|c|c|}
\hline
Parameter  & Value($\mu$m)             \\     
\hline
Patch Length ($L_p$)	&36        \\ 
\hline
Patch Width ($W_p$)	&60\\
\hline
Ring Thickness	&5\\
\hline
Inset Feed Width ($W_1$)	&1\\
\hline
Inset Gap ($W_2$)	&5\\
\hline
Patch cut Width ($W_3$)&4\\
\hline
Patch cut Length ($l_4$)	&10\\
\hline
patch slot Length ($L_c$)	&20\\
\hline
Patch Slot Width ($W_c$) & 5\\
\hline
Substrate Length ($L_s$) & 130\\
\hline
Substrate Width ($W_s$) & 100\\
\hline
Substrate Thickness ($h$) & 45\\
\hline
Microstrip Length ($l_1$) & 40\\
\hline
Inset feed line Length ($l_2$) & 25\\
\hline
Inset cut Length ($l_3$) & 15 \\
\hline
\end{tabular}
\end{table}

The patch antenna is fed by a microstrip feed, which is also made of graphene material. An inset feeding technique is
used to match the impedance of the antenna to the transmission line. Inset feeding involves cutting through the
patch with the transmission line by a particular distance to improve impedance matching, as shown in Fig. \ref{fig:1}. The proposed antenna dimensions are presented in Table \ref{tab1}.

\subsection{Partially Reflecting Sheet Array (PRS)}
\begin{figure}  
		\centering
		\subfloat[Front View ]{\includegraphics[width=0.28\textwidth]{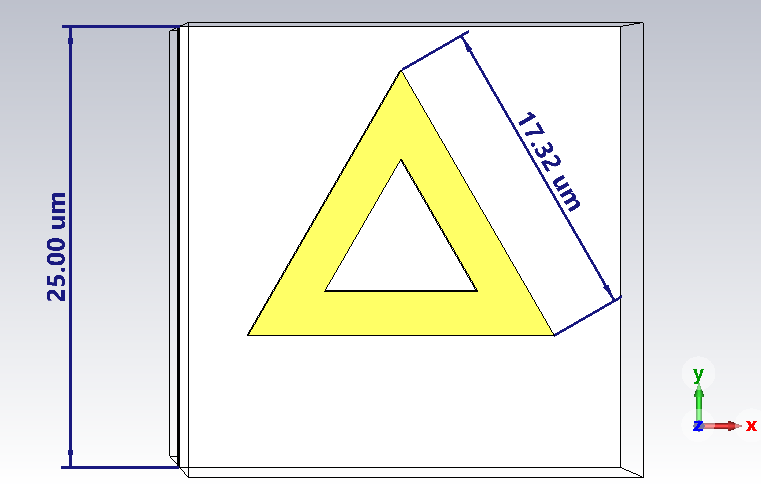} }
		\\
		\subfloat[Back view]{\includegraphics[width=0.28\textwidth]{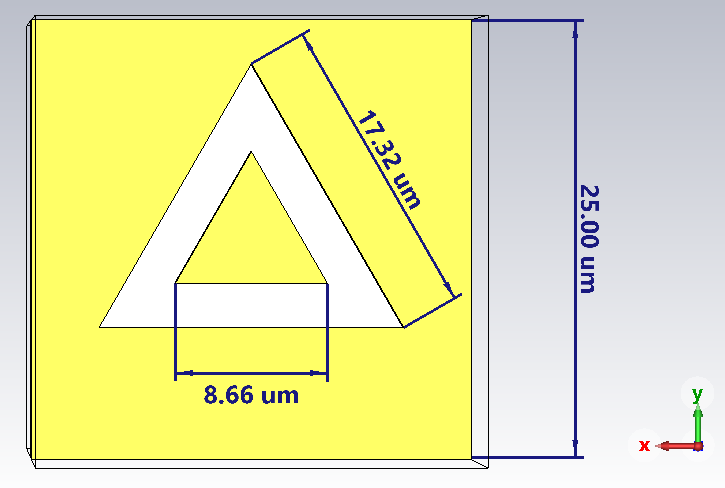}  }\\
        \subfloat[Side view]{\includegraphics[width=0.28\textwidth]{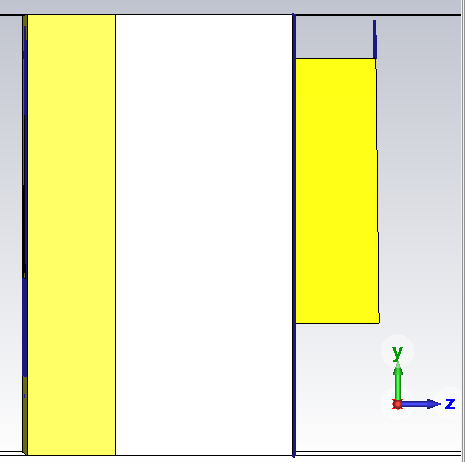}  }\\
		\caption{Proposed unit cell of the PRS.}
		\label{fig:2}
	\end{figure}

A single unit cell is first designed and analysed,
illustrated in Fig. \ref{fig:2}. The unit cell follows a triangular ring structure based on annealed copper, mounted on a
Rogers RT-Duroid 5880 substrate. The substrate has a relative permittivity \begin{math}(\varepsilon_r)\end{math}  2.2 and loss tangent \begin{math}(tan \delta )\end{math} 0.0009 with thickness 10 $\mu $m. Rogers RT-Duroid 5880 dielectric is chosen, as it has a low loss tangent, which implies that the electromagnetic waves travelling through it will experience minimum attenuation. The annealed copper material has a thickness $t_1$ of 5 $\mu$m. The designed
unit cell is much smaller than the patch antenna, as shown in Fig. \ref{fig:2} and Table \ref{tab2}, which implies that the unit cells can be transformed into a physically miniaturised array. This array is later used to improve the gain of the patch antenna due to its unique structure. This structure also increases the antenna's bandwidth due to the triangular ring while experiencing good resonance. 

	\begin{table}
\center
\setlength{\tabcolsep}{1.2em}
\renewcommand{\arraystretch}{1.8}
\caption{dimensions of the designed unit cell} \label{tab2}
\begin{tabular}{|c|c|}
\hline
Parameter  & Value($\mu$m)             \\     
\hline
Ring Outer
Length ($L_g$)	& 17.32      \\ 
\hline
Ring Inner Length ($W_g$)	& 8.66\\
\hline
Ring and Ground
Plane Thickness ($t_1$)	& 5\\
\hline
Substrate
Length ($L_{s_{1}}$)	&1\\
\hline
Substrate
Thickness ($h_1$)	&1\\
\hline
\end{tabular}
\end{table}

\begin{figure}  
		\centering
		\subfloat[PRS array ]{\includegraphics[width=0.28\textwidth]{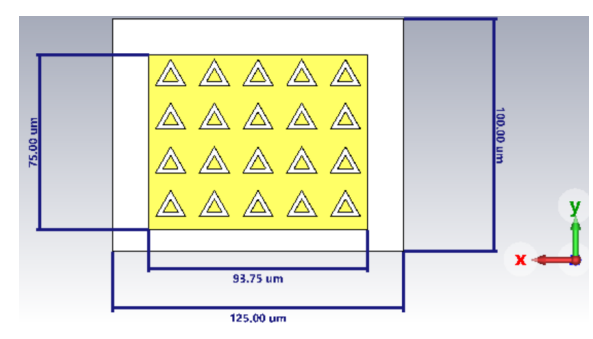} }
		\\
		\subfloat[PRS-stacked
graphene patch antenna]{\includegraphics[width=0.28\textwidth]{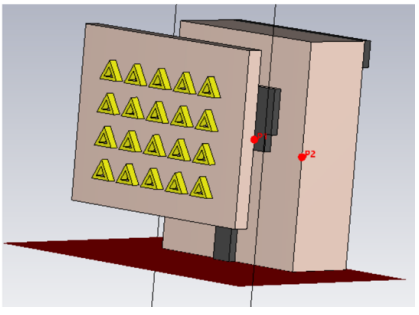}  }\\
		\caption{The PRS sheet array model and the PRS loaded patch antenna.}
		\label{fig:3}
	\end{figure} 
    
\subsection{PRS-Loaded Graphene Patch Antenna}
The fully designed antenna is a PRS metasurface array of 5×4 unit cells positioned above the graphene patch antenna,
as shown in Fig. \ref{fig:3}. The unit cell illustrated in Fig. \ref{fig:2} is used in the array with the same ring
structure and dimensions as shown in Table II. The PRS array and the PRS-stacked graphene patch antenna are shown in
Fig. \ref{fig:3}. The array of unit cells is positioned on a Rogers RT-droid 5880 substrate with thickness 10 $\mu$m.
These unit cells are placed on a rectangular aperture of length $l_{s2}$ = 125 $\mu$m and $W_{s2}$ = 100 $\mu$m. This array of unit cells is considered a periodic structure, where the spacing between adjacent unit cells is in fractions of the wavelength $\lambda$. In this design, the neighbouring unit cells are separated by a distance $\lambda/2$ with respect to each other. Based on this
design, the PRS sheet array is a superstrate structure, enhancing the patch antenna's gain. When a ray
of electromagnetic waves transmitted by the graphene antenna approaches the PRS, it is partly transmitted, and
a part of it gets reflected between the PRS and the antenna several times. Based on the structure of the metasurface,
this is like a capacitor’s structure, as there are radiating elements (the unit cells), a ground plane and a substrate in
between them. This means that when the graphene antenna radiates electromagnetic waves, and as they are incident
on the metasurface, this results in a phase delay to these waves, as they will be stored for a period. Hence, this
corresponds to multiple rays of the same amplitude with equal phase differences. As a result, these rays
experience constructive interference with each other, producing a collimated beam.

    
	
	
	
	

\section{RESULTS AND DISCUSSION}

\begin{figure}  
		\centering
		\subfloat[Return loss (dB) ]{\includegraphics[width=0.5\textwidth]{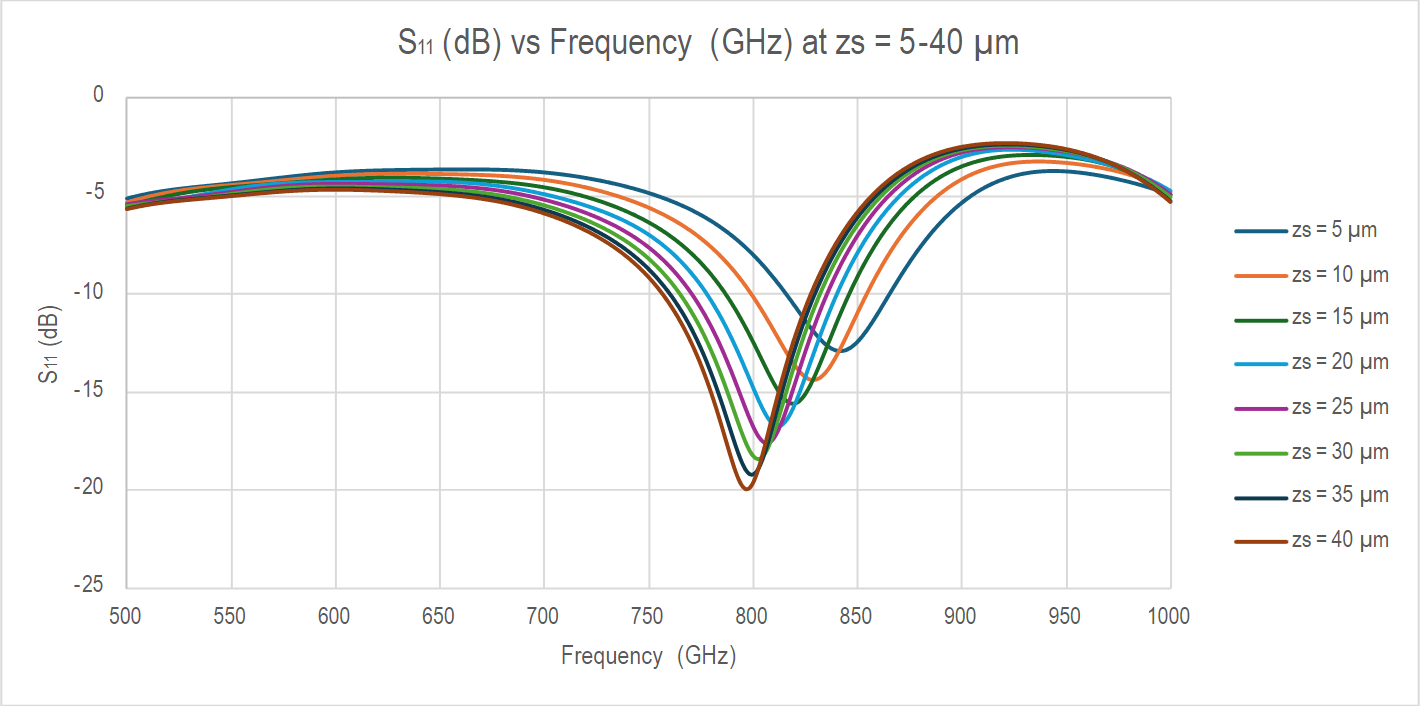} }
		\\
        \subfloat[Reflection phase (deg)]{\includegraphics[width=0.5\textwidth]{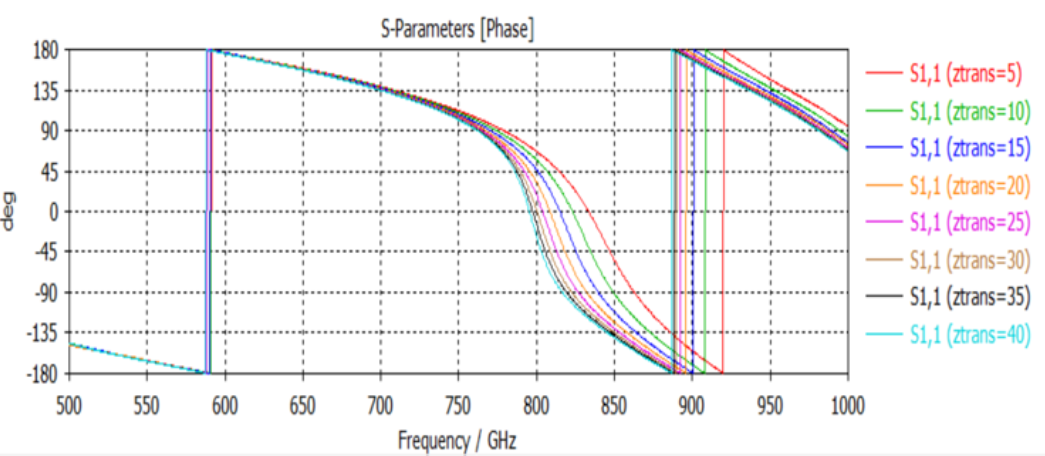}  }\\
		\caption{Reflection Coefficients at $z_s=5\mu m$ to $40\mu m$.}
		\label{fig:4}
	\end{figure}

Fig. \ref{fig:4} illustrates the simulated return loss (dB) and reflection phase of the PRS at different distances between the antenna and the PRS, as the gain improvement from the PRS is highly dependent on this distance. A parametric study is performed for distances $z_s=5\mu m$ to $40\mu m$, with the step size of 5$\mu m$ to examine the optimum resonance distance between the antenna and the PRS. This resulted in an optimised distance of 15$\mu m$, which leads to a gain of 3.6 dBi. The reflection coefficients have been examined at each $z_s$ value, where the magnitude and phase are plotted in Fig. \ref{fig:4}. Based on these s-parameter magnitudes, $z_s$= 40$\mu m$ yielded the highest resonance at ~800 GHz, while $z_s$ = 5 $\mu m$ had the weakest resonance with a resonant frequency of 840 GHz. At $z_s$ =5$\mu m$, this corresponded to the
narrowest bandwidth of 50 GHz. At $z_s$ = 40$\mu m$, the bandwidth equals 70 GHz, which is a significant difference. Fig. \ref{fig:4} (a) shows that the resonance frequency shifts to the left as the distance $z_s$ increases. In addition, the return loss reduces as $z_s$ increases, which implies fewer
reflections of power. At the optimal distance $z_s$ = 15$\mu m$, the resonance frequency equals 820 GHz, with
a -10 dB bandwidth of 65 GHz. As the antenna's bandwidth increases, the dynamic reflection phase curve
becomes more linear. The S11 magnitude plot also conveys how many losses occur in the structure, making it a crucial performance measure. Based on the reflection phase plots in Fig. \ref{fig:4} (b), the resonances occur at the zero
crossing of the vertical axis, which indicates the operating frequency of the antenna; at $z_s$ = 15$\mu m$, the zero-crossing resonance occurs at 820 GHz. Hence, the S11 reflection magnitudes and phase can be used to determine the
resonance frequency. In addition, the transition from a negative phase to a positive phase implies that the behaviour of the PRS changes from capacitive to inductive. The capacitive and inductive behaviour infers that energy is being stored, which isn’t the objective of antennas. At resonance, the behaviour of the PRS is resistive since the reflection phase is equal to zero degrees, where energy is being dissipated.

\begin{figure}  
		\centering
		{\includegraphics[width=0.46\textwidth]{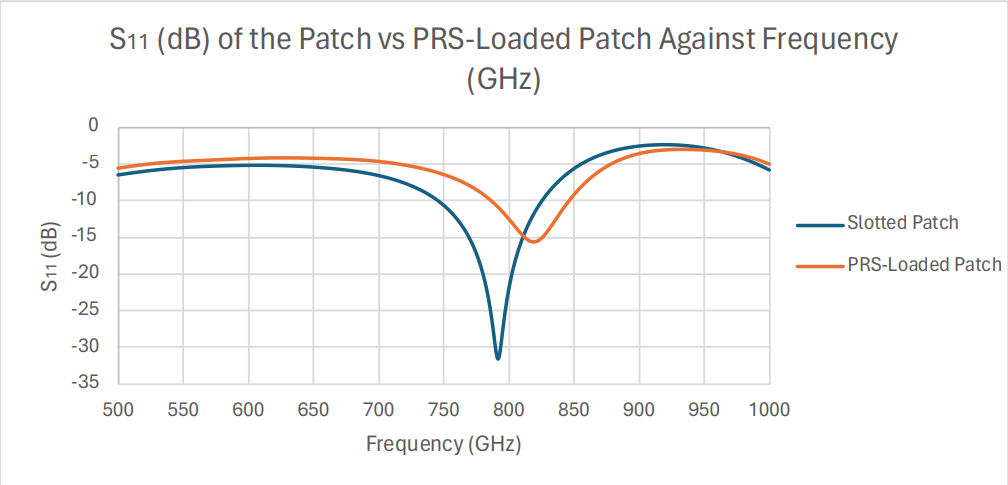} }
		\caption{Return loss (dB) of the slotted patch and PRS-loaded patch.}
		\label{fig:5}
	\end{figure}

    The return loss of the slotted graphene patch is compared with that of the PRS-loaded patch in Fig. \ref{fig:5}.
Based on these results, the PRS shifted the resonance frequency of the slotted patch by 30 GHz. Both designs have
the same bandwidth of 70 GHz with stable operation, as the return loss is low. The
slotted patch has a sharper and stronger resonance at -32 dB. Due to the working principle and structure of the PRS,
which aims at partially reflecting electromagnetic waves, henceforth, the return loss is nearer to 0 dB (fully reflecting
surface).

\begin{figure}  
		\centering
		\subfloat[E-plane ]
		{\includegraphics[width=0.46\textwidth]{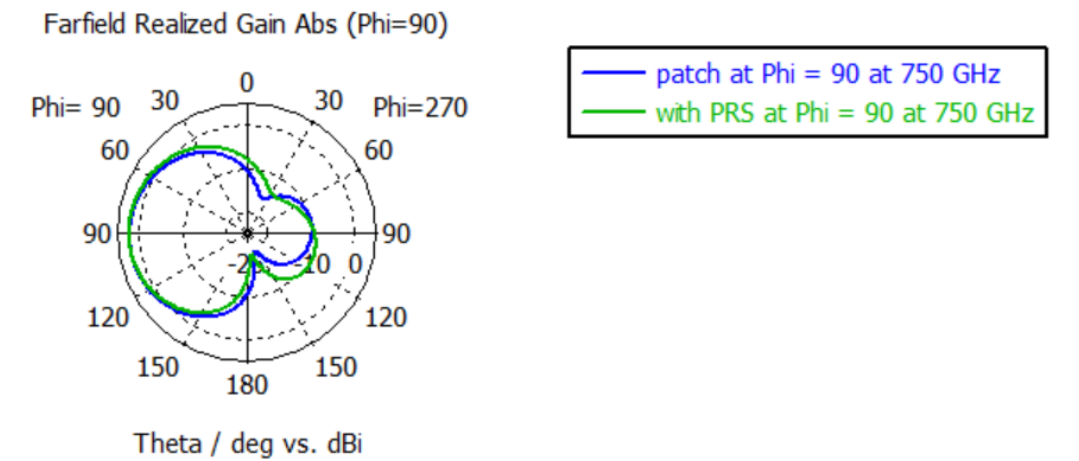} }
		\\
		\subfloat[H-plane]
		{\includegraphics[width=0.46\textwidth]{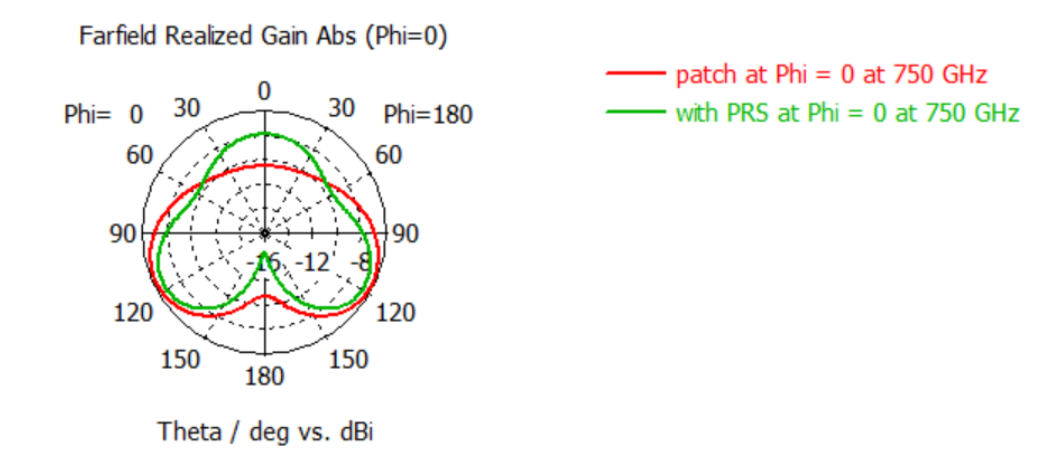}  }\\
		\caption{The farfield results of the slotted patch vs PRS-loaded patch. (a) E-plane. (b) H-plane.}
		\label{fig:6}
	\end{figure}

   The far-field radiation pattern slotted patch is compared with the PRS-loaded patch antenna in Fig. \ref{fig:6}. Both results show stable radiation characteristics. The PRS-based patch has a more significant main lobe level, as shown in
Fig. \ref{fig:6} (a). Both designs have low side lobe magnitudes, which is crucial, as the aim is to have most of the radiated
power directed in the main lobe, depicted by Fig. \ref{fig:6} (a). Based on the H-plane radiation pattern in Fig. \ref{fig:6} (b), the
PRS-loaded patch has a higher main lobe magnitude while having lower side lobe levels. Thus, these results indicate
that the PRS-loaded patch has improved the radiation characteristics of the previously designed slotted patch.

\begin{figure}  
		\centering
		{\includegraphics[width=0.46\textwidth]{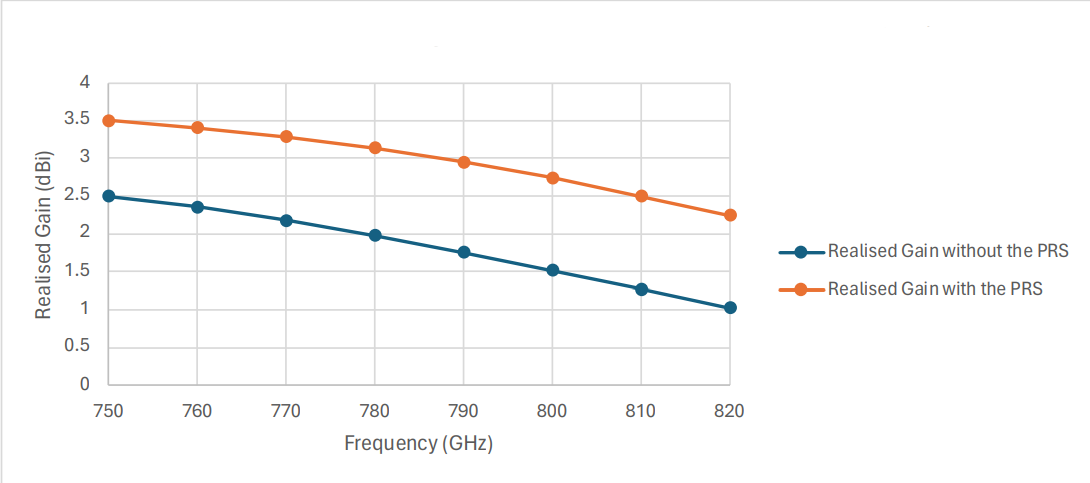} }
		\caption{ Realised gain of the patch vs the PRS-loaded patch.}
		\label{fig:7}
	\end{figure}

Fig. \ref{fig:7} illustrates the realised gain performance of the antenna both with and without the PRS structure as the main objective of incorporating the PRS is to enhance the realised gain of the slotted patch antenna. The results show that the slotted patch antenna has achieved a peak realised gain of 2.49 dBi at 750 GHz. The PRS implementation attained a peak gain of 3.56 dBi. Therefore, introducing a PRS above the antenna enhanced the realised gain by 1.07 dBi. 

\section{Conclusion}
This paper investigates a patch antenna integrated with a PRS for future 6G applications in the THz frequency band. The proposed graphene
antenna attained a 70 GHz (-10 dB) bandwidth. The s-parameters at different scenarios have been
presented and analysed for both the patch and PRS-based antenna. In the future, improvements can be made to the proposed design. For example, the graphene patch antenna can be
transformed into a graphene MIMO antenna, diversifying the antenna design. Another area of further work
could be incorporating the graphene antenna into an array to improve the directivity, gain, and radiation patterns. The designed antenna and PRS can be fabricated on a
PCB and measurements will be taken using a vector network analyser (VNA) to verify the accuracy of the design in
real life, in addition to comparing measurements to simulations.

\bibliographystyle{IEEEtran}
\bibliography{bibliography}
\end{document}